\documentclass[Times,4pt,preprint,onecolumn,superscriptaddress]{revtex4-1}
\usepackage{amsmath,natbib,graphicx,subfigure,amssymb,graphics,amsmath,mathrsfs,CJK,color}
\usepackage{multirow,fancyhdr,color,bm,tabularx,psfrag,geometry,dcolumn,datetime}
\usepackage[colorlinks=true,linkcolor=blue,urlcolor=blue,citecolor=blue]{hyperref}
\usepackage[left]{lineno}%[switch]
\def\footnoterule{\kern -1mm \hrule width 6.6cm \kern 2.2mm}%
\usepackage{tikz,xcolor,hyperref}% Make Orcid icon
\definecolor{lime}{HTML}{A6CE39}
\DeclareRobustCommand{\orcidicon}{%
    \begin{tikzpicture}
    \draw[lime, fill=lime] (0,0)
    circle [radius=0.16] node[white]
   {{\fontfamily{qag}\selectfont \tiny ID}};\draw[white, fill=white] (-0.0625,0.095)
    circle [radius=0.007];
    \end{tikzpicture}
    \hspace{-2mm}}
\foreach \x in {A, ..., Z}
{\expandafter\xdef\csname orcid\x\endcsname{\noexpand\href{https://orcid.org/\csname orcidauthor\x\endcsname}{\noexpand\orcidicon}}}

\begin{document}

\title{\LARGE Negative refraction with little loss manipulated by the voltage and pulsed laser in double quantum dots}

\author{Shun-Cai Zhao\orcidA{}}
\email[Corresponding author: ]{zsczhao@126.com}
\affiliation{Department of Physics, Faculty of Science, Kunming University of Science and Technology, Kunming, 650500, PR China}
\affiliation{Center for Quantum Materials and Computational Condensed Matter Physics, Faculty of Science, Kunming University of Science and Technology, Kunming, 650500, PR China}
\author{Xiao-fan Qian}
\affiliation{Department of Physics, Faculty of Science, Kunming University of Science and Technology, Kunming, 650500, PR China}
\affiliation{Center for Quantum Materials and Computational Condensed Matter Physics, Faculty of Science, Kunming University of Science and Technology, Kunming, 650500, PR China}
\author{Ya-Ping Zhang}
\affiliation{Department of Physics, Faculty of Science, Kunming University of Science and Technology, Kunming, 650500, PR China}
\affiliation{Center for Quantum Materials and Computational Condensed Matter Physics, Faculty of Science, Kunming University of Science and Technology, Kunming, 650500, PR China}
\author{Yong-An Zhang}
\affiliation{Department of Physics, Faculty of Science, Kunming University of Science and Technology, Kunming, 650500, PR China}
\affiliation{Center for Quantum Materials and Computational Condensed Matter Physics, Faculty of Science, Kunming University of Science and Technology, Kunming, 650500, PR China}

%\date{\currenttime,~\today}

\begin{abstract}
The paper demonstrates that negative refractive index can be
achieved via tuning the tunneling rate between a double quantum
dots(QDs) system by applying a bias voltage, and a pulsed laser. As
the bias voltage being changed, the refraction index can be tunable
to negative with the simultaneous negative permittivity and
permeability. While the varying pulsed laser is applied to the
double QDs system, moreover, the negative refractive index with
little loss can be obtained. The flexible manipulation on a solid
state system to realize negative refraction may give a new way for
experimental research in the future.
\begin{description}
\item[PACS: ]{78.20.Ci, 42.50.Gy }
%\item[Keywords]{ Photoexcited carriers transfer, doped double quantum dots photocell, photovoltaic properties}
\end{description}
\end{abstract}

\maketitle
\section{Introduction}
Negative refraction materials\cite{Ref1},which originally suggested
by Veselago attracted impressive efforts to be investigated during
the past decades. Materials with negative refractive index promise
many surprising and even counterintuitive electromagnetical and
optical effects such as the negative Goos-H$\ddot{a}$nchen
shift\cite{Ref2}, amplification of evanescent waves\cite{Ref3},
reversals of both Doppler shift and Cerenkov radiation\cite{Ref2},
sub-wavelength focusing\cite{Ref5} and so on. And there have been
many approaches to realize materials with negative refractive index,
which can be summarized as artificial structures such as
metamaterials \cite{Ref6} and photonic crystals\cite{Ref7}, chiral
materials\cite{Ref8} and photonic resonant media\cite{Ref9}-\cite{Ref10} . 
The scheme of photonic resonant media firstly
brought forth by M.$\ddot{O}$.Oktel\cite{Ref9} discussed how a
three-level atomic gas system to realize optical modification of
magnetic permeability and then possibly to obtain left-handed
electrodynamics. It found that it is in principle possible to
electromagnetically induce left-handedness to a spatially
homogeneous media. The major challenge for the scheme is to have two
levels separated at optical frequencies while having a nonvanishing
magnetic dipole matrix element. Such level splitting requires large
external magnetic fields or should be engineered by other means such
as external electric fields or spin-orbital couplings. Therefore, we
can consider solid state systems\cite{Ref9}, and try to utilize
excitonic energy levels in solid state heterostructures to engineer
three-level system satisfying the energy condition.

Thus, from the point of against the challenge, QDs with a suitable
spectrum may be the candidate. QDs provide a three-dimensional
confinement of carriers,in which electrons and holes can occupy only
a set of states with discrete energies, and can thus be used to
perform ``atomic physics''experiments in solid-state structures. One
advantage of QD¡¯s is that allow the control of direct
quantum-mechanical electronic coupling with not only composition but
externally applied voltages. These flexible systems represent
therefore the ideal for theoretical and experimental investigations,
where the interactions between light and matter can be studied in a
fully controlled, well characterized environment, and with excellent
optical and electrical probes. These features make semiconductor
QD¡¯s promising candidates for applications in electro-optical
devices such as QD lasers\cite{Ref22}-\cite{Ref23} and in quantum
information processing\cite{Ref24}-\cite{Ref25}. In the latter
case, one can exploit the optical excitation in a QD\cite{Ref24},
or its spin state\cite{Ref25}, as qubits. The ability to assemble
collections of QD¡¯s with designed geometries opens up a number of
interesting possibilities.

The aim of this paper is to explore the feasibility that negative
refractive index can be achieved by tuning the tunneling rate
between the double QD system via applying an external bias voltage,
and an optical pump pulse. Via the bias voltage one can suppress or
enhance the tunneling rate between the two dots. As the tunneling
rate is changed, the double quantum dot system is tunable to
left-handedness with simultaneous negative permittivity and
permeability. Varying the intensity of the pulsed laser, the quantum
dot system can also show left-handedness with negative refractive
index. Moreover, the negative refractive index and little loss can
be achieved at the same time.

In the present paper, the model system is introduced in the next
section. In Sec.III, the results of our calculations for two
different conditions are presented by varying the bias voltage and
the intensity of the pulse. Finally, we give a summary of our
results and conclusions in Sec.IV.

\section{Model System}

\begin{figure}
\centerline{
\includegraphics[width=7 cm,height=4 cm]{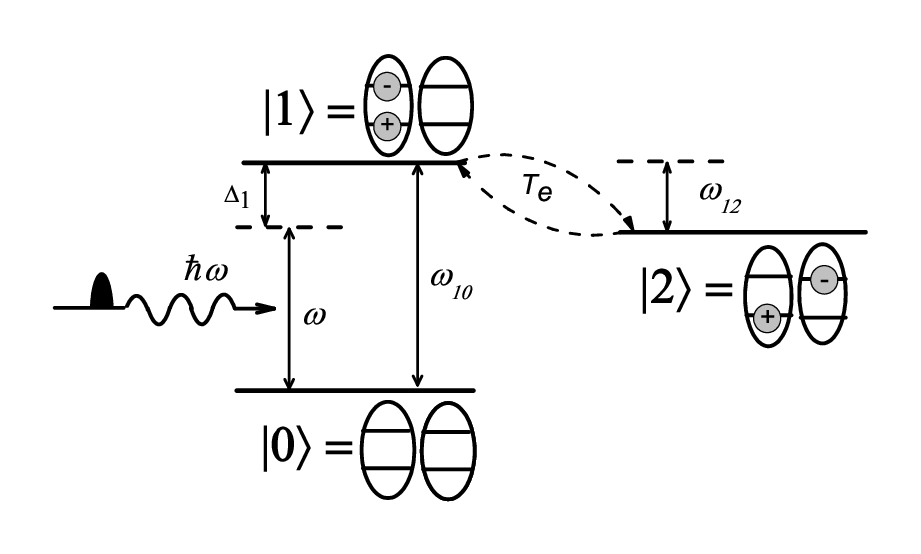}}
\caption{Schematic representation of the relevant atomic energy
levels for the double QDs system. A pulsed laser excites one
electron from the valence band that can tunnel to the other dot. We
assume that the hole cannot be tunneled here.}
\label{fig:1}
\end{figure}

A schematic representation of the Hamiltonian for our model is shown
in Figure 1. The double QDs consists of the left and the right dots
with different band structures coupled by tunneling in our model
system. In the setup, the self-assembled lateral QD molecules were
produced on GaAs(0,0,1) substrates by a unique combination of
molecule beam epitaxy and in situ atomic layer precise
etching\cite{Ref11}-\cite{Ref13} which provides a low density
homogeneous ensembles of QD molecules consisting of two dots aligned
along the [1,$\bar{1}$,0] direction. The lateral quantum coupling
between two self-assembled (In,Ga)As quantum dots has been
observed\cite{Ref12}, and it demonstrated that the QD molecule is
composed of two distinct QDs, even though the QDs are separated by
only a few nanometers of barrier material. And we know that for the
QD's separation d $>$ 9nm the electron tunnel coupling energy
between the two dots is smaller as compared to the electron-hole
Coulomb interaction energy and the exciton states can readily be
described in terms of a simplified single-particle picture\cite{Ref14}-\cite{Ref16}. Because the nanoscale interdot separation, the
exciton states consist of delocalized electron states, i.e., holes
are localized in the QDs, while electrons become almost entirely
delocalized in the QD molecules\cite{Ref12}. By applying
electromagnetic field we can excite one electron from the valence to
the conduction band in the left dot, which can in turn tunnel to the
right one. Applying a bias voltage, we can adjust the electron
tunneling rate from the left dot to the right one. So the tunnel
barrier in the double QDs can be controlled by placing a gate
electrode between the two QDs. In general, there may be several
excitation channels between the two dots. In order to capture the
main physics we do not take into account multi-channel effect here.
Figure 1 also depicts the energy-level diagram for the double QDs
system. The ground state $|0\rangle$ is the system without
excitations, and the exciton state $|1\rangle$ describes a pair of
electron and hole bound in the left dot. The indirect exciton state
$|2\rangle$ represents one hole in the left dot with an electron in
the right dot. The two energy levels of $|0\rangle$ and $|1\rangle$
have opposite parity with
$\vec{d_{10}}=\langle1|$$\hat{\vec{d}}$$|0\rangle$$\neq0$ , where
$\hat{\vec{d}}$ is the electric dipole operator. While the levels
$|0\rangle$ and $|2\rangle$ have the same parity and so
$\vec{\mu_{20}}=\langle2|$$\hat{\vec{\mu}}$$|0\rangle$$\neq0$.
$\hat{\vec{\mu}}$ is the magnetic-dipole operator. Using this
configuration the effective Hamiltonian of the system reads as
follows\cite{Ref 17}(with $\hbar=1$):
\begin{equation}
H=\frac{1}{2}\left(
                   \begin{array}{ccc}
                     -\Delta_{1} & 2\Omega & 0 \\
                     2\Omega & \Delta_{1} & 2Te \\
                     0 &  2Te &3\Delta_{1}-2 \Delta_{2} \\
                   \end{array}
                 \right)
\end{equation}
where $\Delta_{1}=\omega_{10}-\omega$ is the detuning of the pulsed
laser with the exciton resonance transition
$|0\rangle\leftrightarrow|1\rangle$. Here $\Omega =\langle0|\mu
\cdot\textbf{E}(t)|1\rangle/2\hbar$ is the Rabi frequency of the
pulsed laser(with the angular frequency $\omega$) to drive the
transition $|0\rangle\leftrightarrow |1\rangle$, where the electric
dipole moment $\vec{\mu}$ describes the coupling to the radiation
field of the excitonic transition. And $\textbf{E}(t)$ is the
electric field amplitude of the pulsed laser. $T_{e}$ is the
electron tunneling rate\cite{Ref17}. The parameters $T_{e}$ and
$\omega_{12}$ can be tuned by the bias voltage. The detuning
$\Delta_{2}$ is defined as $\Delta_{2}=\Delta_{1}-\omega_{12}$,
where $\omega_{ij}=\omega_{i}-\omega_{j}$, with the energies of the
$|i\rangle$ and $|j\rangle$ states being $\hbar\omega_{i}$ and
$\hbar\omega_{j}$. Using the density-matrix approach, the
time-evolution of the system is described as
$\frac{d\rho}{dt}=-\frac{i}{\hbar}[H,\rho]+\Lambda\rho$, where
$\Lambda\rho$ represents the irreversible decay part in the system.
Here, $\Lambda\rho$ is a phenomenological added decay term that
corresponds to the incoherent processes. Using this configuration
the dynamics of the system can be described by the following density
matrix equations,
\begin{eqnarray}
\dot{\rho_{01}}=i(\triangle_{1}+i\gamma_{1})\rho_{01}-i\Omega(\rho_{11}-\rho_{00})+iT_{e}\rho_{02}\\
\dot{\rho_{12}}=-i(\triangle_{1}-\triangle_{2}-i\gamma_{2})\rho_{12}-i\Omega\rho_{02}-iT_{e}(\rho_{22}-\rho_{11})\\
\dot{\rho_{02}}=i(\triangle_{2}+i\gamma_{3})\rho_{02}-i\Omega\rho_{12}+iT_{e}\rho_{01}\\
\dot{\rho_{00}}=\Gamma_{20}\rho_{22}+\Gamma_{10}\rho_{11}-i\Omega(\rho_{10}-\rho_{01})\\
\dot{\rho_{11}}=-(\Gamma_{10}+\Gamma_{12})\rho_{11}+i\Omega(\rho_{10}-\rho_{01})-iT_{e}(\rho_{21}-\rho_{12})
\end{eqnarray}
with $\rho_{00}+\rho_{11}+\rho_{22}=1$ , $\rho_{ij}$=
$\rho_{ji}^{\ast }$ , and with $i\neq j$, i , j = 0, 1, 2 .
 Where $\Gamma_{ij}$ denotes the decay rate of the populations from state $|i\rangle$ to state $|j\rangle$
, and  $\gamma_{1}, \gamma_{2}, \gamma_{3}$ depict the decay rates
of coherence of the off-diagonal density matrix element for
$\rho_{10},\rho_{12}$ and $\rho_{20}$, respectively.

According to the classical electromagnetic theory,the electric
polarizability is a rank 2 tensor defined by its Fourier transform
$\vec{P}_{e}(\omega_{P})$$=\epsilon_{0}$$\alpha_{e}(\omega_{P})$$\vec{E}(\omega_{P})$,
which is calculated as the mean value of the atomic electric-dipole
moment operators by the definition $\vec{P}_{e}$
=Tr$\{$${\hat{\rho}\vec{d}}$$\}$=$\rho_{01}d_{10}$+c.c. where Tr
stands for trace. In the following, we only consider the
polarizability at the frequency $\omega_{P}$ of the incoming field
$\vec{E}_{p}$. Therefore we drop the explicit $\omega_{P}$
dependence $\alpha_{e}(\omega_{P})\equiv\alpha_{e}$. Moreover, we
choose $\vec{E}_{p}$ parallel to the atomic dipole $\vec{d}_{10}$ so
that $\alpha_{e}$is a scalar, and its expression is as follows:

\begin{equation}
\alpha_{e}=\frac{\vec{d}_{10}\rho_{01}}{\epsilon_{0}\vec{E}_{p}}=\frac{\mid
{d_{10}}\mid^{2} \rho_{01}}{\epsilon_{0}\hbar\Omega},
\end{equation}

In the same way, the classical  magnetic polarizations of the medium
$\vec{P}_{m}(\omega_{P})$=$\mu_{0}$ $\alpha_{m}\vec{E}(\omega_{P})$,
which is related to the mean value of the atomic dipole moment
operator through $\vec{P}_{m}$
=Tr$\{$${\hat{\rho}\vec{\mu}}$$\}$=$\rho_{02}$$\mu_{20}+c.c$.
According to the classical Maxwell's electromagnetic wavevector
relation,we choose magnetic dipole is perpendicular to the induced
electric dipole so that the magnetizability $\alpha_{m}$ is scalar,
and its expression is as follows:
\begin{equation}
\alpha_{m}=\frac{\mu_{0}\vec{\mu}_{20}\rho_{02}}{\vec{B}_{p}},
\end{equation}
with the relation between the microscopic local electric and
magnetic fields $\vec{E}_{p}/\vec{B}_{p}= c$, we can obtain the
explicit expression for the atomic magnetic polarizability
$\alpha_{m}$.

According to the Clausius-Mossotti relations and considering the
local effect in dense medium\cite{Ref18}, the relative permittivity
and relative permeability are read as\cite{Ref19}
\begin{eqnarray}
\epsilon_{r}=\frac{1+\frac{2}{3}N\alpha_{e}}{1-\frac{1}{3}N\alpha_{e}},
\end{eqnarray}
\begin{eqnarray}
\mu_{r}=\frac{1+\frac{2}{3}N\alpha_{m}}{1-\frac{1}{3}N\alpha_{m}}.
\end{eqnarray}

In the above, we obtained the expressions for the electric
permittivity and magnetic permeability of the double quantum dots
system. In the section that follows, we will demonstrate that the
simultaneously negative both permittivity and permeability, negative
refraction index with little absorption can be observed in the QDs
system.

\section{The analysis of results}

In the following,with the stationary solutions to the density-matrix
equations(2)-(6), we explore the sign property of  both electric
permittivity and magnetic permeability through the numerical
calculations. In what follows we choose the parameters for the QDs
system, $d_{10}$=2.5$\times$ $10^{-29}$C$\cdot$m and $\mu_{20}$=7.0
$\times$$10^{-23}$ C$\cdot m^{2}s^{-1}$\cite{Ref20}. We choose the
density of atoms N to be $5\times10^{21}m^{-3}$,
$\gamma=10^{7}s^{-1}$. For simplify,the other parameters are scaled
by $\gamma$: $\Gamma_{10}=\Gamma_{12}=\Gamma_{20}=0.5\gamma$,
$\gamma_{1}=\gamma_{2}=0.1\gamma$, $\gamma_{3}=0.25\gamma$. The Rabi
frequency of the pulsed laser is $\Omega=0.1\gamma$. Figure 2 shows
the calculated electric permittivity $\epsilon_{r}$ and magnetic
permeability $\mu_{r}$ as a function of the tunneling rate $T_{e}$
with the pulsed laser coupling the levels $|0\rangle$ and
$|1\rangle$ resonantly. And the detuning $\Delta_{2}$ varis with
three different values: $0.1\gamma$,  $0.3\gamma$ and $0.5\gamma$.

\begin{figure}
\centerline{
\includegraphics[width=6 cm,height=4 cm]{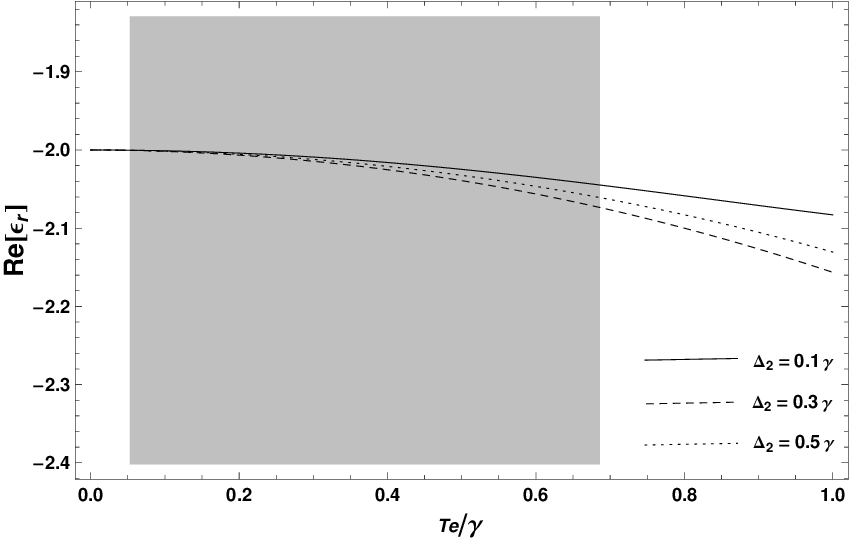}
\includegraphics[width=6 cm,height=4 cm]{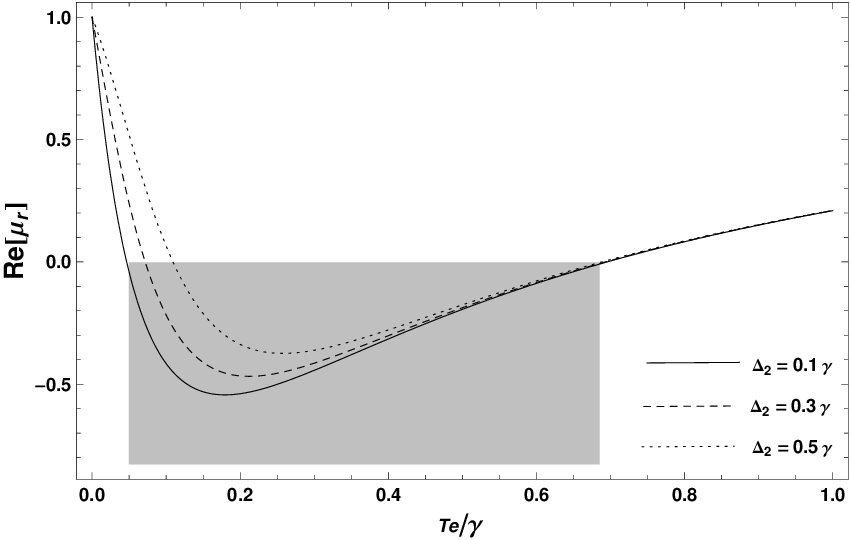}
}
\end{figure}
\begin{figure}
\centerline{
\includegraphics[width=6 cm,height=4 cm]{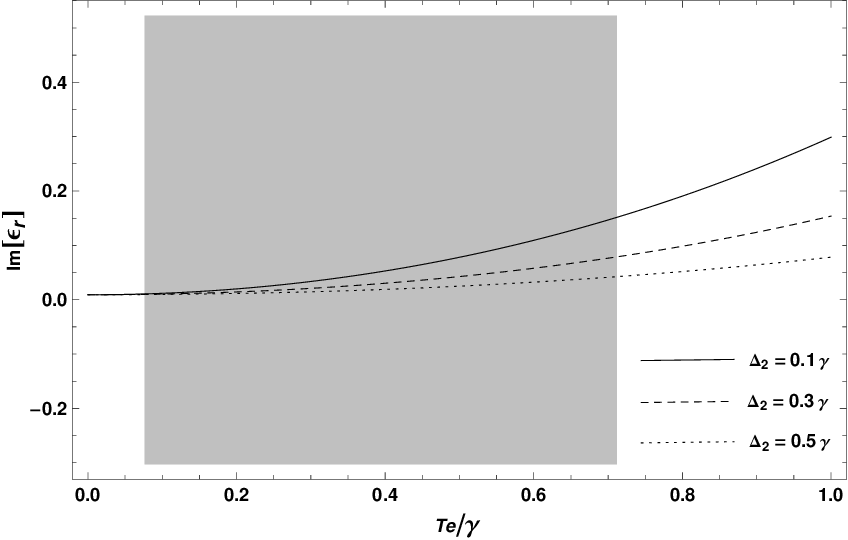}
\includegraphics[width=6 cm,height=4 cm]{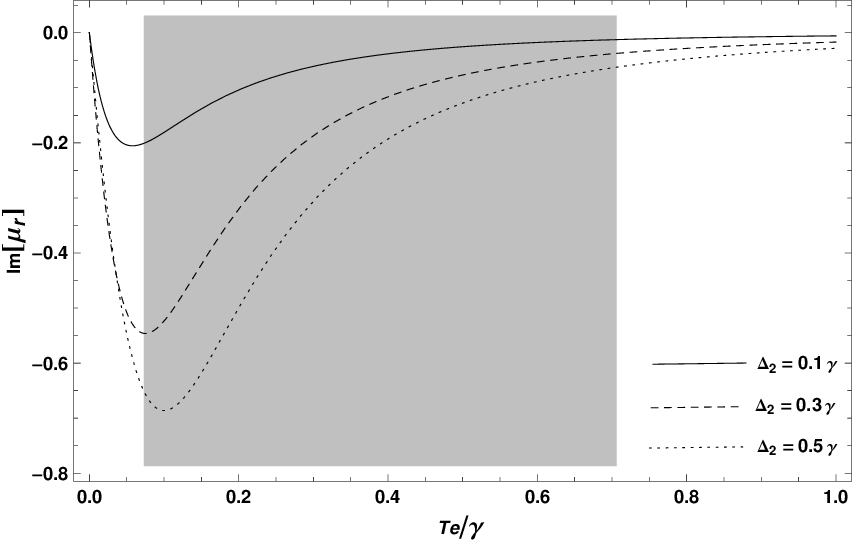}}
%\vspace{0cm}
\caption{The real and imaginary parts of the permittivity
$\varepsilon_{r}$ and the permeability $\mu_{r}$
  as a function of the tunneling rate $Te$ with different
  detunings $\Delta_{2}$. The other parameters are given in the text.}
\label{fig:2}
\end{figure}

For the parameters we choose, left-handed properties of the QDs
system are obtained inside the shaded area of Figure 2. We notice
that the bandwidth for left-handed properties decreases with the
increasing of the detuning $\Delta_{2}$, which is displayed by the
computed dependence of the real permeability on the tunneling rate
$T_{e}$. And $\Delta_{2}=0.1\gamma$ with the largest bandwidth of
about 6.3 MHz, $\Delta_{2}=0.5\gamma$ with the narrowest bandwidth
of about 5.5 MHz. The QDs system is passive medium for the
permittivity because of its real and imaginary parts having opposite
sign, and increasing active for the permeability which imaginary
parts have minus sign and increasing value.

\begin{figure}
\centerline{
\includegraphics[width=6 cm,height=4 cm]{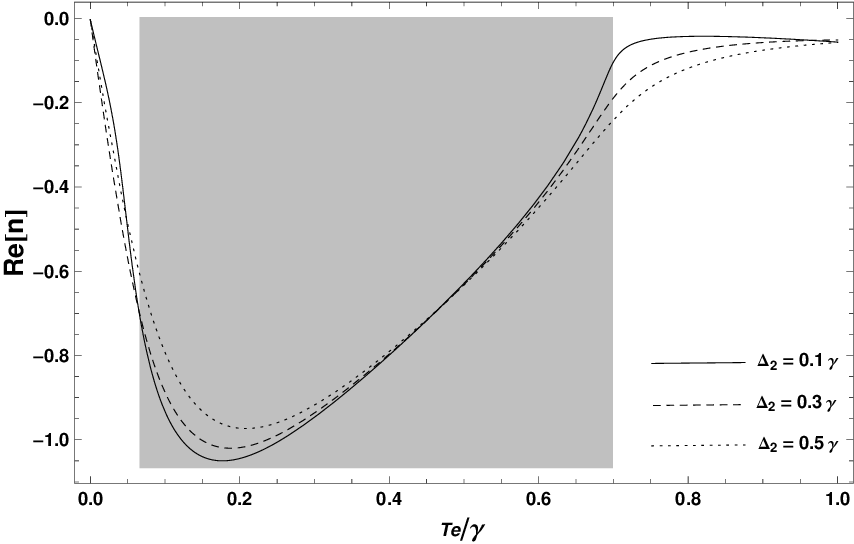}
\includegraphics[width=6 cm,height=4 cm]{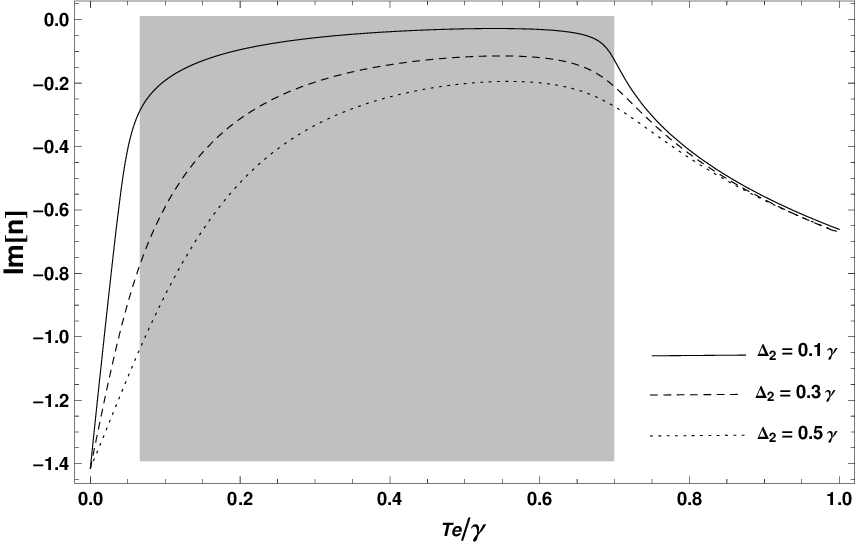} }
%\vspace{5cm}
\caption{The refractive index as a function of the tunneling rate
$Te$ with different detunings $\Delta_{2}$. The other parameters are
same as in figure 2.} \label{fig:3}
\end{figure}

According to the refraction definition of the left-handed material
$(n(\omega)=-\sqrt{\epsilon_{r}(\omega) \mu_{r}(\omega)})$, Figure 3
presents the computed dependence of the refraction index on the
tunneling rate $T_{e}$. In the shaded area, the real part of the
refraction index shows the shrinking values and the imaginary part
displays increasing gain response.

Above, the bias voltage manipulating the tunneling rate $T_{e}$ on
the QDs system's left-handed properties is taken into account.
However, another one, the effect of pulsed laser should be included.
Figure 4 depicts the dependence of the permittivity and permeability
on the coupling intensity of the pulsed laser. And the tunneling
rate $T_{e}=0.2\gamma$, the detuning $\Delta_{2}$ is varied by
$0.03\gamma$, $0.05\gamma$, $0.1\gamma$. The other parameters are
the same to the former. The left-handed properties of the QDs system
can be provided inside the shaded area varying the values of
$\Delta_{2}$.

\begin{figure}
\centerline{%
\includegraphics[width=6 cm,height=4 cm]{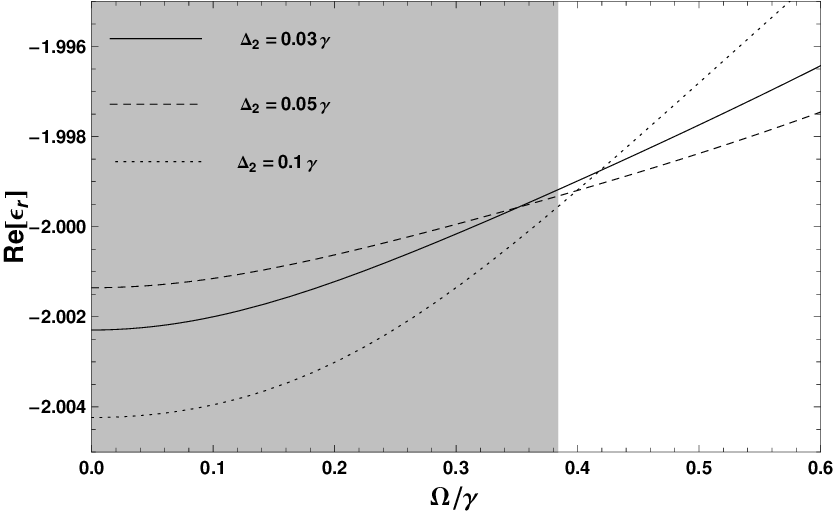}
\includegraphics[width=6 cm,height=4 cm]{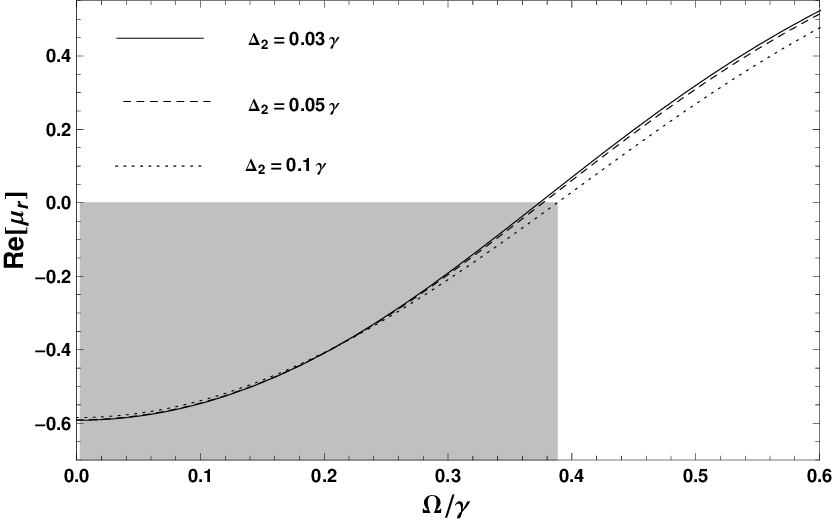}}
\end{figure}
\begin{figure}
\centerline{%
\includegraphics[width=6 cm,height=4 cm]{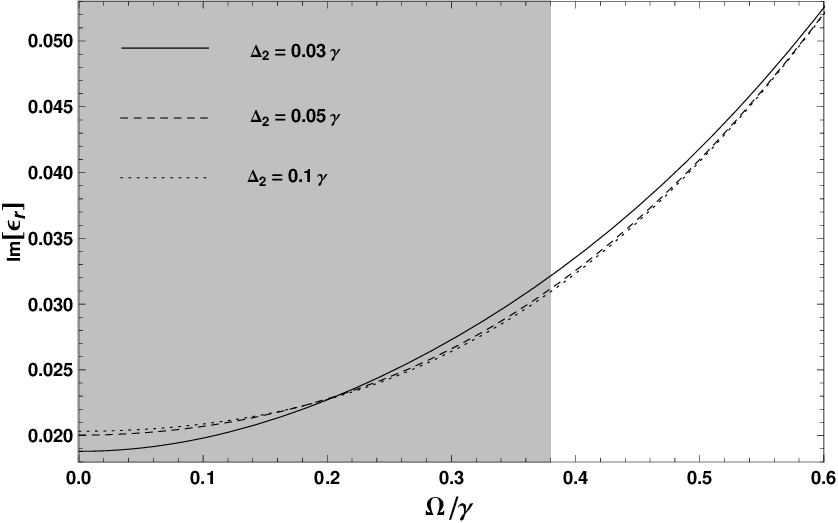}
\includegraphics[width=6 cm,height=4 cm]{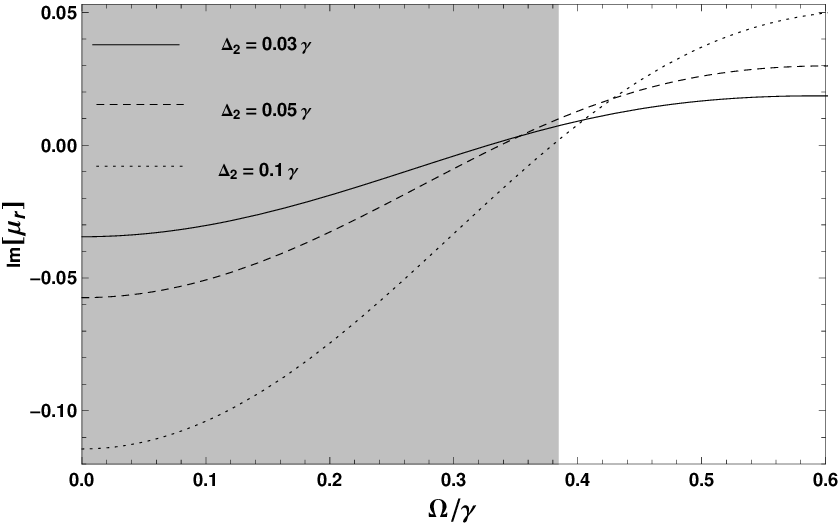}}
\caption{The real and imaginary parts of the permittivity
$\varepsilon_{r}$ and the permeability $\mu_{r}$
  as a function of the intensity of the pulsed laser$\Omega$ with with different
  detunings $\Delta_{2}$. The other parameters are given in the text.}
\label{fig:4}
\end{figure}

\begin{figure}
\centerline{%
\includegraphics[width=6 cm,height=4 cm]{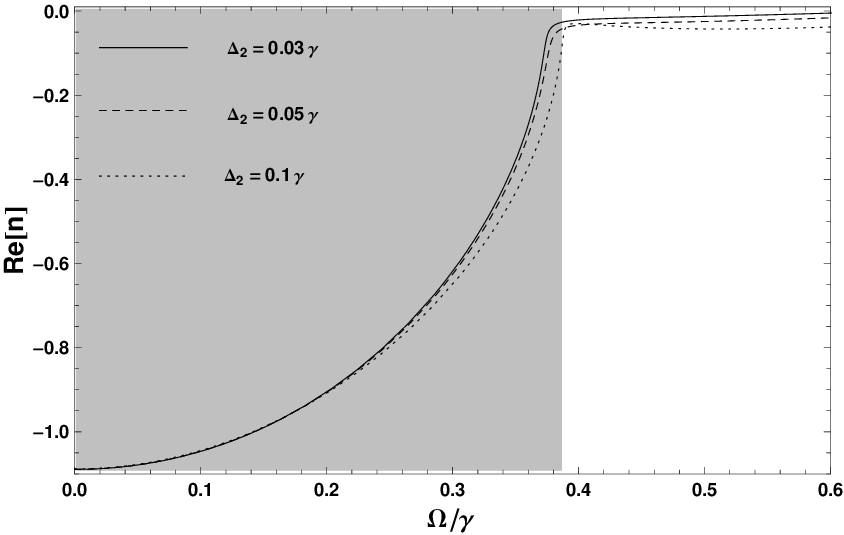}
\includegraphics[width=6 cm,height=4 cm]{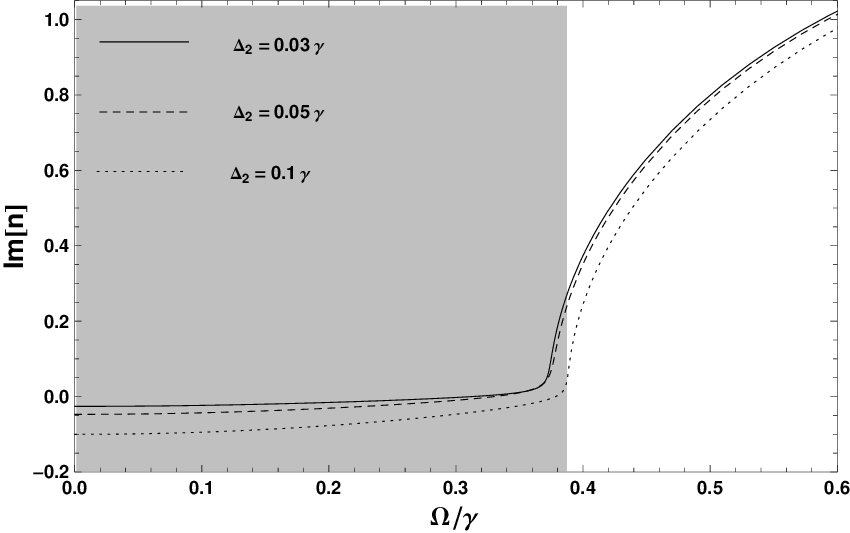}}
\caption{The refractive index as a function of the intensity of the
pulsed laser $\Omega$ with different detunings $\Delta_{2}$. The
other parameters are same as in figure 4.}
\label{fig:5}
\end{figure}

As observed in Figure 5, the real part of the refractive index shows
negative values decreasing gradually with the increasing of
$\Omega$. The imaginary part of the refractive index displays gain
decreasing and near the transparency.

\begin{figure}
\centerline{%
\includegraphics[width=6 cm,height=4 cm]{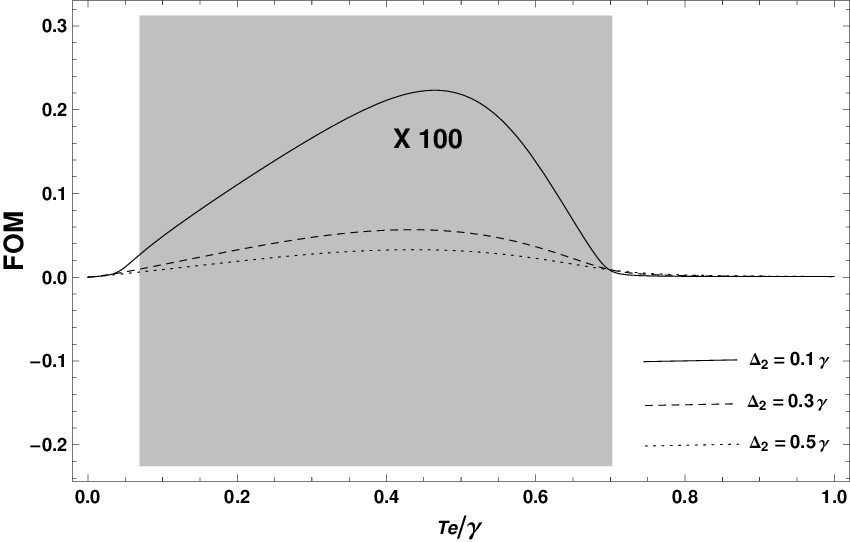}
\includegraphics[width=6 cm,height=4 cm]{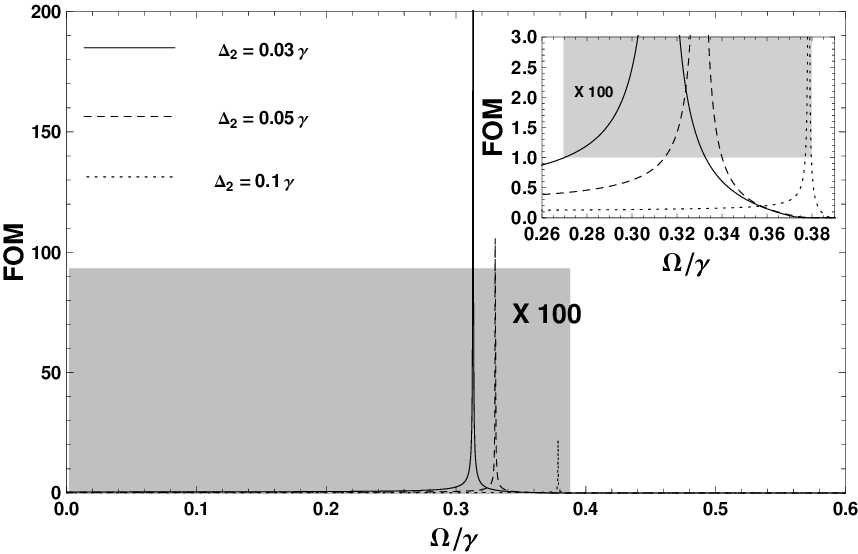}}
\caption{Figure of merit ($FOM:|real(n)/imag(n)|$) as a function of
the tunneling rate $Te$(in the left) and the intensity of the laser
pulse $\Omega$(in the right).}
\label{fig:6}
\end{figure}

The figure of merit(FOM) $|real(n)/imag(n)|$ \cite{Ref21} of
negative index materials which has to be seriously considered since
the low loss of negative refractive index materials is desired. When
the FOM is much larger than unity, it means that there is almost
little loss in this area. Via manipulating the bias voltage and the
pulsed laser intensity, Figure 6 shows the figure of merit(FOM) the
double QDs system in the shaded areas. We notice that the FOM is
decreasing and far less than unity in the shaded area with different
$\Delta_{2}$ values when tunneled the tunneling rate $T_{e}$via the
bias voltage in the left of Figure 6. This illustrates that the QDs
system displays left-handedness but the increasing loss when
applying the bias voltage to tunnel the tunneling rate $T_{e}$.
However, varying the pulsed laser intensity the QDs system shows the
different results. In the right one of Figure 6, the FOM has three
peak and much larger than unity when varied $\Delta_{2}$ by
$0.03\gamma$, $0.05\gamma$ and $0.1\gamma$. The inset in the right
shows the peaks of FOM present proximately at
$\Omega=0.316\gamma$,$0.325\gamma$ and $0.38\gamma$. It means that
there is little loss for the QDs system when we applied the pulsed
laser to the QDs system. The markedly feature of present scheme is
the flexible manipulation on a solid state system to achieve the
negative refractive index. We demonstrate left-handedness in the
double QDs system applying the bias voltage and the pulsed
laser.What's more, when the changing pulsed laser is applied to the
QDs system, the left-handedness with little loss can also be
obtained.

\section{The conclusions}

In conclusion, we have demonstrated that left-handedness with
negative refraction properties can be realized in a solid system,
the double QDs system. It shows that negative refractive index with
simultaneous negative permittivity and permeability can be achieved
by tuning the tunneling rate between the double quantum dots via
applying a bias voltage. Moreover, the negative refractive index
with little loss can be obtained when the varying pulsed laser is
applied to the double QDs system. This will be very helpful to the
potential applications in optical devices designed. The varied
tunneling rate via the bias voltage and the pulsed laser applied to
the double QDs system give a more flexible manipulation on a solid
system to realized negative refraction in the coming experimental
research.

\section*{Acknowledgements}
The work is supported by the National Natural Science Foundation of
China (Grant Nos.61168001)and the Foundation for Personnel training
projects of Yunnan Province,China (Grant No.KKSY201207068).

%\appendix
%\section{First Appendix} %Empty argument \section{} yields `Appendix'.
%
%\section{Second Appendix}


\begin{thebibliography}{0}%
\makeatletter
\providecommand \@ifxundefined [1]{%
 \@ifx{#1\undefined}
}%
\providecommand \@ifnum [1]{%
 \ifnum #1\expandafter \@firstoftwo
 \else \expandafter \@secondoftwo
 \fi
}%
\providecommand \@ifx [1]{%
 \ifx #1\expandafter \@firstoftwo
 \else \expandafter \@secondoftwo
 \fi
}%
\providecommand \natexlab [1]{#1}%
\providecommand \enquote  [1]{``#1''}%
\providecommand \bibnamefont  [1]{#1}%
\providecommand \bibfnamefont [1]{#1}%
\providecommand \citenamefont [1]{#1}%
\providecommand \href@noop [0]{\@secondoftwo}%
\providecommand \href [0]{\begingroup \@sanitize@url \@href}%
\providecommand \@href[1]{\@@startlink{#1}\@@href}%
\providecommand \@@href[1]{\endgroup#1\@@endlink}%
\providecommand \@sanitize@url [0]{\catcode `\\12\catcode `\$12\catcode
  `\&12\catcode `\#12\catcode `\^12\catcode `\_12\catcode `\%12\relax}%
\providecommand \@@startlink[1]{}%
\providecommand \@@endlink[0]{}%
\providecommand \url  [0]{\begingroup\@sanitize@url \@url }%
\providecommand \@url [1]{\endgroup\@href {#1}{\urlprefix }}%
\providecommand \urlprefix  [0]{URL }%
\providecommand \Eprint [0]{\href }%
\providecommand \doibase [0]{http://dx.doi.org/}%
\providecommand \selectlanguage [0]{\@gobble}%
\providecommand \bibinfo  [0]{\@secondoftwo}%
\providecommand \bibfield  [0]{\@secondoftwo}%
\providecommand \translation [1]{[#1]}%
\providecommand \BibitemOpen [0]{}%
\providecommand \bibitemStop [0]{}%
\providecommand \bibitemNoStop [0]{.\EOS\space}%
\providecommand \EOS [0]{\spacefactor3000\relax}%
\providecommand \BibitemShut  [1]{\csname bibitem#1\endcsname}%
\let\auto@bib@innerbib\@empty
%</preamble>
\end{thebibliography}%


\begin{thebibliography}{99}
%%%%%%%%%%%%%%%%%%%%%%%%%%%%%%%%%%%%%%%%%%%%%%%%%%%%%%%%%%%%%
% Some macros are available for the bibliography:
%  o for general use
%    \JL : general journals                 \andvol : Vol (Year) Page
%  o for individual journal
%    \AJ   : Astrophys. J.           \NC         : Nuovo Cim.
%    \ANN  : Ann. of Phys.           \NPA, \NPB  : Nucl. Phys. [A,B]
%    \CMP  : Commun. Math. Phys.     \PLA, \PLB  : Phys. Lett. [A,B]
%    \IJMP : Int. J. Mod. Phys.      \PRA - \PRE : Phys. Rev. [A-E]
%    \JHEP : J. High Energy Phys.    \PRL        : Phys. Rev. Lett.
%    \JMP  : J. Math. Phys.          \PRP        : Phys. Rep.
%    \JP   : J. of Phys.             \PTP        : Prog. Theor. Phys.
%    \JPSJ : J. Phys. Soc. Jpn.      \PTPS       : Prog. Theor. Phys. Suppl.
% Usage:
%  \PRD{45,1990,345}          ==> Phys.~Rev.\ D \textbf{45} (1990), 345
%  \JL{Nature,418,2002,123}   ==> Nature \textbf{418} (2002), 123
%  \andvol{123,1995,1020}    ==> \textbf{123} (1995), 1020
%%%%%%%%%%%%%%%%%%%%%%%%%%%%%%%%%%%%%%%%%%%%%%%%%%%%%%%%%%%%%

\bibitem{Ref1}V.G.Veselago and E.E.Narimanov, {\it Nature Mater.}, \textbf{5},(2006),759.

\bibitem{Ref2}P.R.Berman, {\it Phys. Rev. E} {\bf 66},(2002),067603.

\bibitem{Ref3}M.W.Feise, P.J.Bevelacqua and J.B.Schneider, {\it Phys.Rev.B},{\bf 66},(2002), 035113.

\bibitem{Ref5}K.Aydin, I.Bulu and E.Ozbay, {\it Appl. Phys. Lett} {\it 90},(2007)254102.

\bibitem{Ref6}R.A.Shelby and D.R.Smith, {\it Science}, {\bf 77},(2001)292.

\bibitem{Ref7}E.Cubukcu, {\it Nature}, {\bf 423},(2003),604.

\bibitem{Ref8}J.K$\ddot{a}$tel, M.Fleischhauer, S.F.Yelin and P.L.Walsworth,{\it Phys.Rev.Lett.} {\bf 99},(2007), 073602.

\bibitem{Ref9}M. $\ddot{O}$. Oktel, $\ddot{O}$. E. M$\ddot{u}$tecapl$\check{g}$u, {\it Phys. Rev. A},{\bf 70},(2004),053806.

\bibitem{Ref22}O.B. Shchekin, G. Park, D.L. Huffaker, and D.G. Deppe, {\it Appl. Phys. Lett. }, {\bf 77},(2000),466.

\bibitem{Ref23}S.C. Zhao, Z.D. Liu, J. Zheng, G. Li, N. Liu,{\it Optik },{\bf 123},(2012),1063.

\bibitem{Ref24}A. Ekert and R. Jozsa, {\it Rev. Mod. Phys.}, {\bf 68},(1996),733.

\bibitem{Ref25}A. Imamo$\breve{g}$lu, D.D. Awschalom, G. Burkard, D.P. DiVincenzo, D. Loss, M. Sherwin, and A. Small, {\it Phys. Rev. Lett.}, {\bf 83},(1999),4204.

\bibitem{Ref10}S.C.Zhao, Z.D.Liu and Q.X.Wu, {\it J. Phys. B}, {\bf 43 },(2010),045505.

\bibitem{Ref11}B.Krause and T.H.Metzger, {\it Phys. Rev. B}, {\bf 72},(2005),085339.

\bibitem{Ref12}G.J.Beirne, C.Hermannst$\ddot{a}$ter, L.Wang, A.Rastelli, O.G.Schmidt and P.Michler, {\it Phys. Rev. Lett.}, {\bf 96},(2006),137401.

\bibitem{Ref13}R.Songmuang, S.Kiravittaya and O.G.Schmidt, {\it Appl.Phys.Lett.}, {\bf 82},(2003,.2892 .

\bibitem{Ref14}H.J.Krenner, M.Sabathil, E.C.Clark, A. Kress, D. Schuh, M. Bichler, G. Abstreiter, J.J. Finley, A. Zrenner, {\it Phys. Rev. Lett.}, {\bf 94},(2005),057402.

\bibitem{Ref15}H.J. Krenner, S. Stufler, M. Sabathil, E.C.Clark, P. Ester, M. Bichler, G. Abstreiter, J.J. Finley, A. Zrenner, {\it New J.Phys.},{\bf 7},(2005),184.

\bibitem{Ref16}G.Bester, A.Zunger, J.Shumway, {\it Phys. Rev. B}, {\bf 71},(2005),075325.

\bibitem{Ref17}J. M. Villas-B$\hat{o}$s, A .O. Govorov and Sergio E. Ulloa, {\it Phys. Rev. B}, {\bf 69},(2004), 125342.

\bibitem{Ref18}G. S. Agarwal, R.W.Boyd,{\it Phys. Rev. A }, {\bf 60},(1999),R2681.

\bibitem{Ref19}J.Q.Shen,{\it J. Mod. Opt.}, {\bf 53},(2006),2195-2205.

\bibitem{Ref20}Q. Thommen, P. Mandel, {\it Phys. Rev. Lett.}{\bf 96},(2006),053601.

\bibitem{Ref21}J.K$\ddot{a}$tel and M. Fleischhauer, {\it Phys. Rev. A }, {\bf 79},(2009),063818.

\end{thebibliography}
\end{document}